\documentclass[a4paper,11pt]{article}
\pdfoutput=1
\usepackage{jinstpub}
\notoc
\usepackage{lineno}   % line numbers are mandatory for TWEPP26
\usepackage[british]{babel}
\usepackage{siunitx} %
\usepackage[nohyperlinks]{acronym}
\usepackage{xcolor}
\usepackage{subcaption}
\usepackage{float}
\usepackage{booktabs}
\usepackage{tikz}
\usetikzlibrary{arrows.meta, positioning, calc}

\definecolor{figureInk}{RGB}{35,35,35}
\definecolor{figureBorder}{RGB}{100,100,100}
\definecolor{figureFill}{RGB}{246,246,246}
\definecolor{schem}{RGB}{49,73,94}
\definecolor{figureAccentFill}{RGB}{236,240,243}

\newcommand{\PowerChainFigure}{%
  \resizebox{\linewidth}{!}{%
  \begin{tikzpicture}[
      font=\small, text=figureInk,
      stage/.style={draw=figureBorder, rounded corners, minimum width=3cm, minimum height=1.4cm,
                    align=center, fill=white},
      card/.style={draw=figureBorder, rounded corners, minimum width=2.7cm, minimum height=0.95cm,
                   align=center, fill=white, font=\footnotesize},
      s3box/.style={draw=figureBorder, rounded corners, minimum width=3.8cm, minimum height=2.7cm,
                    fill=figureFill},
      iface/.style={draw=schem, rounded corners, fill=figureAccentFill, text=schem,
                    font=\footnotesize\bfseries, inner sep=3pt},
      stagelbl/.style={text=figureInk, font=\footnotesize\bfseries,
                       anchor=north, inner sep=0pt},
      arrow/.style={-{Stealth[length=2.5mm]}, thick, draw=figureInk},
      ifline/.style={schem, thick},
      clen/.style={font=\footnotesize\itshape, text=figureInk},
  ]
      \node[stage] (s1) {\textbf{ELTEK Flatpack2}\\[2pt]\qty{400}{\volt}\,\acs*{AC} $\rightarrow$ \qty{380}{\volt}\,\acs*{DC}};
      \node[stage, right=2.3cm of s1] (s2) {\textbf{MaraPOD}\\[2pt]\qty{380}{\volt} $\rightarrow$ \qty{12}{\volt}};
      \node[s3box, right=2.3cm of s2] (s3) {};
      \node[card] (c1) at ($(s3.center)+(0.3,0.6)$) {\textbf{\acs*{LVR} card 1}\\$4\times$ bPOL12V};
      \node[card] (c2) at ($(s3.center)+(0.3,-0.6)$) {\textbf{\acs*{LVR} card 2}\\$4\times$ bPOL12V};
      \coordinate (in) at ($(s1.west)+(-1.5,0)$);
      \draw[arrow] (in) -- (s1.west);
      \node[above, font=\footnotesize] at ($(in)!0.32!(s1.west)$) {\qty{400}{\volt}\,\acs*{AC}};
      \draw[arrow] (s1.east) -- (s2.west);
      \coordinate (m1) at ($(s1.east)!0.5!(s2.west)$);
      \fill[schem] (m1) circle (1.6pt);
      \node[clen] at ($(m1)+(0,-0.42)$) {\qtyrange[range-phrase=--,range-units=single]{80}{200}{\metre}};
      \node[iface] (if1) at ($(m1)+(0,1.75)$) {Interface 1};
      \draw[ifline] (if1.south) -- (m1);
      \draw[arrow] (s2.east) -- (s3.west);
      \coordinate (m2) at ($(s2.east)!0.5!(s3.west)$);
      \fill[schem] (m2) circle (1.6pt);
      \node[clen] at ($(m2)+(0,-0.42)$) {\qtyrange[range-phrase=--,range-units=single]{20}{40}{\metre}};
      \node[iface] (if2) at ($(m2)+(0,1.75)$) {Interface 2};
      \draw[ifline] (if2.south) -- (m2);
      \coordinate (bjoin) at ($(s3.west)+(0.4,0)$);
      \draw[thick] (s3.west) -- (bjoin);
      \draw[thick] ($(bjoin)+(0,0.6)$) -- ($(bjoin)+(0,-0.6)$);
      \draw[arrow] ($(bjoin)+(0,0.6)$) -- (c1.west);
      \draw[arrow] ($(bjoin)+(0,-0.6)$) -- (c2.west);
      \coordinate (lbly) at ($(s3.south)+(0,-0.15)$);
      \node[stagelbl] at (s1 |- lbly) {Stage 1};
      \node[stagelbl] at (s2 |- lbly) {Stage 2};
      \node[stagelbl] at (s3 |- lbly) {Stage 3};
  \end{tikzpicture}%
  }%
}

\newcommand{\MeasSetupFigure}{%
  \resizebox{\linewidth}{!}{%
  \begin{tikzpicture}[
      >={Latex[length=1.5mm]}, font=\footnotesize, text=figureInk,
      box/.style={draw=figureBorder, rounded corners=2pt, align=center,
        inner sep=3pt, fill=white},
      power/.style={semithick, draw=figureInk},
      injection/.style={semithick, draw=schem},
      sense/.style={line width=0.45pt, dash pattern=on 2.5pt off 1.8pt,
        draw=figureBorder},
      dot/.style={circle, fill=figureInk, inner sep=1.2pt},
    ]
    \node[box, minimum width=1.5cm, minimum height=1.2cm] (source) at (0.8,0.2)
      {\textbf{MaraPOD}\\\qty{12}{\volt}};
    \node[box, fill=figureAccentFill, minimum width=1.15cm,
      minimum height=1.2cm] (load) at (5.8,0.2) {\textbf{\acs*{LVR}}\\load};
    \draw[power] (source.east |- 0,0.55) -- (load.west |- 0,0.55);
    \draw[power] (source.east |- 0,-0.15) -- (load.west |- 0,-0.15);

    \fill[white] (2.6,0.55) circle (0.19);
    \draw[injection] (2.6,0.55) circle (0.19);
    \draw[injection] (2.6,0.55) circle (0.10);
    \node[anchor=south, inner sep=1pt] at (2.6,0.77) {Current probe};

    \node[box, draw=schem, fill=figureAccentFill,
      minimum width=1.5cm, minimum height=0.7cm] (dc) at (3.4,1.8) {\acs*{DC} block};
    \node[box, minimum width=1.8cm, minimum height=0.85cm] (inj) at (1.2,1.8)
      {B-WIT~100\\injector};
    \draw[injection] (inj.east |- 0,1.95) -- (dc.west |- 0,1.95);
    \draw[injection] (dc.east |- 0,1.95) -- (4.5,1.95) -- (4.5,0.55);
    \draw[injection] (inj.east |- 0,1.55) -- (2.3,1.55) -- (2.3,1.30) -- (4.1,1.30)
      -- (4.1,0.64) arc(90:-90:0.09) -- (4.1,-0.15);
    \node[dot] at (4.5,0.55) {};
    \node[dot] at (4.1,-0.15) {};

    \node[box, fill=figureFill, minimum width=1.8cm, minimum height=1.1cm]
      (bode) at (0.9,-1.35) {\textbf{Bode~100}\\analyser};
    \coordinate (drivePort) at (bode.west |- 0,-1.0);
    \coordinate (currentPort) at (bode.east |- 0,-1.0);
    \coordinate (voltagePort) at (bode.east |- 0,-1.45);
    \draw[sense, ->] (drivePort) -- (-0.65,-1.0)
      -- (-0.65,1.8) -- (inj.west);
    \node[anchor=west, fill=white, inner sep=1pt] at (-0.65,1.0) {\acs*{AC} drive};

    \draw[sense, ->] (2.6,0.36) -- (2.6,-0.06)
      arc(90:-90:0.09) -- (2.6,-1.0) -- (currentPort);
    \node[anchor=south, inner sep=2pt] at ($(currentPort)!0.5!(2.6,-1.0)$) {$I$};

    \node[box, minimum width=2cm, minimum height=0.85cm] (diff) at (4.6,-1.45)
      {Differential\\amplifier};
    \coordinate (diffPlus) at ($(diff.north)+(-0.45,0)$);
    \coordinate (diffMinus) at ($(diff.north)+(0.45,0)$);
    \draw[sense] (4.5,0.55) -- (4.5,-0.06) arc(90:-90:0.09)
      -- (4.5,-0.65) -- (diffPlus |- 0,-0.65) -- (diffPlus);
    \draw[sense] (diffMinus |- 0,-0.15) -- (diffMinus);
    \node[dot] at (diffMinus |- 0,-0.15) {};
    \node[anchor=east, inner sep=0pt] at ($(diffPlus)+(-0.10,0.14)$) {$+$};
    \node[anchor=west, inner sep=0pt] at ($(diffMinus)+(0.10,0.14)$) {$-$};
    \draw[sense, ->] (diff.west) -- (voltagePort);
    \node[anchor=south, inner sep=2pt] at ($(voltagePort)!0.5!(diff.west)$) {$V$};
  \end{tikzpicture}%
  }%
}

\providecommand{\ValidationPanelLabel}{}
\providecommand{\ValidationPanelLabel}{}
\usepackage{pgfplots}
\usepgfplotslibrary{groupplots}
\pgfplotsset{compat=1.18}
\definecolor{impedanceLight}{RGB}{100,100,100}
\definecolor{impedanceHeavy}{RGB}{155,76,49}

\newcommand{\ImpedanceOverlayFigure}{%
  \begin{tikzpicture}
    \begin{groupplot}[
        group style={group size=1 by 2, vertical sep=0.15cm,
          x descriptions at=edge bottom},
        width=0.72\linewidth, scale only axis,
        xmode=log, xmin=100, xmax=1e7,
        xtick={1e2,1e3,1e4,1e5,1e6,1e7},
        xmajorgrids, ymajorgrids,
        grid style={draw=figureBorder!25, very thin},
        axis line style={draw=figureBorder},
        tick style={draw=figureBorder},
        tick label style={font=\footnotesize, text=figureInk},
        label style={font=\footnotesize, text=figureInk},
        every axis plot/.append style={line width=0.7pt, no marks},
      ]
      \nextgroupplot[
        height=1.10cm, ymode=log, ymin=0.3, ymax=200,
        ytick={1,10,100}, ylabel={$|\Zin|$ (\unit{\ohm})},
      ]
        \path[fill=figureBorder!8] (axis cs:1e6,0.3) rectangle (axis cs:1e7,200);
        \addplot[impedanceLight, forget plot] table[col sep=space,x=f,y=mag]
          {figures/data/impedance/s2_zin_meas_N1.dat};
        \addplot[impedanceLight, densely dashed, forget plot] table[col sep=space,x=f,y=mag]
          {figures/data/impedance/s2_zin_sim_N1.dat};
        \addplot[schem, forget plot] table[col sep=space,x=f,y=mag]
          {figures/data/impedance/s2_zin_meas_N6.dat};
        \addplot[schem, densely dashed, forget plot] table[col sep=space,x=f,y=mag]
          {figures/data/impedance/s2_zin_sim_N6.dat};
        \addplot[impedanceHeavy, forget plot] table[col sep=space,x=f,y=mag]
          {figures/data/impedance/s2_zin_meas_N12.dat};
        \addplot[impedanceHeavy, densely dashed, forget plot] table[col sep=space,x=f,y=mag]
          {figures/data/impedance/s2_zin_sim_N12.dat};
      \nextgroupplot[
        height=0.90cm, ymin=-145, ymax=100, ytick={-90,0,90},
        ylabel={$\angle\Zin$ (\unit{\degree})},
        xlabel={Frequency (\unit{\hertz})},
        legend to name=impedanceLoadLegend, legend columns=4,
        legend style={draw=none, font=\footnotesize,
          /tikz/every even column/.append style={column sep=5pt}},
      ]
        \path[fill=figureBorder!8] (axis cs:1e6,-145) rectangle (axis cs:1e7,100);
        \addplot[impedanceLight, forget plot] table[col sep=space,x=f,y=phase]
          {figures/data/impedance/s2_zin_meas_N1.dat};
        \addplot[impedanceLight, densely dashed, forget plot] table[col sep=space,x=f,y=phase]
          {figures/data/impedance/s2_zin_sim_N1.dat};
        \addplot[schem, forget plot] table[col sep=space,x=f,y=phase]
          {figures/data/impedance/s2_zin_meas_N6.dat};
        \addplot[schem, densely dashed, forget plot] table[col sep=space,x=f,y=phase]
          {figures/data/impedance/s2_zin_sim_N6.dat};
        \addplot[impedanceHeavy, forget plot] table[col sep=space,x=f,y=phase]
          {figures/data/impedance/s2_zin_meas_N12.dat};
        \addplot[impedanceHeavy, densely dashed, forget plot] table[col sep=space,x=f,y=phase]
          {figures/data/impedance/s2_zin_sim_N12.dat};
        \addlegendimage{empty legend}
        \addlegendentry{$N_{\mathrm{T}}$:}
        \addlegendimage{impedanceLight, line width=0.7pt}
        \addlegendentry{$1$}
        \addlegendimage{schem, line width=0.7pt}
        \addlegendentry{$6$}
        \addlegendimage{impedanceHeavy, line width=0.7pt}
        \addlegendentry{$12$}
    \end{groupplot}
    \node[anchor=north west,font=\footnotesize,fill=white,inner sep=1pt]
      at (group c1r1.north west) {\ValidationPanelLabel};
  \end{tikzpicture}%
  \par\vspace{-3pt}
  \pgfplotslegendfromname{impedanceLoadLegend}%
}

\definecolor{validationMid}{RGB}{114,67,111}
\pgfplotsset{
  validation axis/.style={
    width=0.72\linewidth, scale only axis,
    xmode=log, xmin=100, xmax=1e7,
    xtick={1e2,1e3,1e4,1e5,1e6,1e7},
    xmajorgrids, ymajorgrids,
    grid style={draw=figureBorder!25, very thin},
    axis line style={draw=figureBorder}, tick style={draw=figureBorder},
    tick label style={font=\footnotesize, text=figureInk},
    label style={font=\footnotesize, text=figureInk},
    every axis plot/.append style={line width=0.7pt, no marks},
  },
}

\newcommand{\EltekImpedanceValidationFigure}{%
  \begin{tikzpicture}
    \begin{groupplot}[
      validation axis,
      group style={group size=1 by 2, vertical sep=0.15cm,
        x descriptions at=edge bottom},
    ]
      \nextgroupplot[
        height=1.10cm, ymode=log, ymin=1e-3, ymax=20,
        ytick={1e-3,1e-1,1e1}, ylabel={$|\Zout|$ (\unit{\ohm})},
      ]
        \path[fill=figureBorder!8] (axis cs:1e6,1e-3) rectangle (axis cs:1e7,20);
        \addplot[forget plot,impedanceLight] table[col sep=space,x=f,y=mag]
          {figures/data/impedance/s1_zout_meas_N2.dat};
        \addplot[forget plot,impedanceLight,densely dashed] table[col sep=space,x=f,y=mag]
          {figures/data/impedance/s1_zout_sim_N2.dat};
        \addplot[forget plot,schem] table[col sep=space,x=f,y=mag]
          {figures/data/impedance/s1_zout_meas_N6.dat};
        \addplot[forget plot,schem,densely dashed] table[col sep=space,x=f,y=mag]
          {figures/data/impedance/s1_zout_sim_N6.dat};
        \addplot[forget plot,validationMid] table[col sep=space,x=f,y=mag]
          {figures/data/impedance/s1_zout_meas_N12.dat};
        \addplot[forget plot,validationMid,densely dashed] table[col sep=space,x=f,y=mag]
          {figures/data/impedance/s1_zout_sim_N12.dat};
        \addplot[forget plot,impedanceHeavy] table[col sep=space,x=f,y=mag]
          {figures/data/impedance/s1_zout_meas_N16.dat};
        \addplot[forget plot,impedanceHeavy,densely dashed] table[col sep=space,x=f,y=mag]
          {figures/data/impedance/s1_zout_sim_N16.dat};
      \nextgroupplot[
        height=0.90cm, ymin=-105, ymax=160, ytick={-90,0,90},
        ylabel={$\angle\Zout$ (\unit{\degree})}, xlabel={Frequency (\unit{\hertz})},
        legend to name=eltekValidationLegend, legend columns=5,
        legend style={draw=none, font=\footnotesize,
          /tikz/every even column/.append style={column sep=3pt}},
      ]
        \path[fill=figureBorder!8] (axis cs:1e6,-105) rectangle (axis cs:1e7,160);
        \addplot[forget plot,impedanceLight] table[col sep=space,x=f,y=phase]
          {figures/data/impedance/s1_zout_meas_N2.dat};
        \addplot[forget plot,impedanceLight,densely dashed] table[col sep=space,x=f,y=phase]
          {figures/data/impedance/s1_zout_sim_N2.dat};
        \addplot[forget plot,schem] table[col sep=space,x=f,y=phase]
          {figures/data/impedance/s1_zout_meas_N6.dat};
        \addplot[forget plot,schem,densely dashed] table[col sep=space,x=f,y=phase]
          {figures/data/impedance/s1_zout_sim_N6.dat};
        \addplot[forget plot,validationMid] table[col sep=space,x=f,y=phase]
          {figures/data/impedance/s1_zout_meas_N12.dat};
        \addplot[forget plot,validationMid,densely dashed] table[col sep=space,x=f,y=phase]
          {figures/data/impedance/s1_zout_sim_N12.dat};
        \addplot[forget plot,impedanceHeavy] table[col sep=space,x=f,y=phase]
          {figures/data/impedance/s1_zout_meas_N16.dat};
        \addplot[forget plot,impedanceHeavy,densely dashed] table[col sep=space,x=f,y=phase]
          {figures/data/impedance/s1_zout_sim_N16.dat};
        \addlegendimage{empty legend}\addlegendentry{$N_{\mathrm{T}}$:}
        \addlegendimage{impedanceLight}\addlegendentry{$2$}
        \addlegendimage{schem}\addlegendentry{$6$}
        \addlegendimage{validationMid}\addlegendentry{$12$}
        \addlegendimage{impedanceHeavy}\addlegendentry{$16$}
    \end{groupplot}
    \node[anchor=north west,font=\footnotesize,fill=white,inner sep=1pt]
      at (group c1r1.north west) {\ValidationPanelLabel};
  \end{tikzpicture}%
  \par\vspace{-3pt}
  \pgfplotslegendfromname{eltekValidationLegend}%
}

\newcommand{\SingleTerminalValidationFigure}[5]{%
  \begin{tikzpicture}
    \begin{groupplot}[
      validation axis,
      group style={group size=1 by 2, vertical sep=0.15cm,
        x descriptions at=edge bottom},
    ]
      \nextgroupplot[
        height=1.10cm, ymode=log, ymin=#3, ymax=#4,
        ytick={#5}, ylabel={$|#2|$ (\unit{\ohm})},
      ]
        \path[fill=figureBorder!8] (axis cs:1e6,#3) rectangle (axis cs:1e7,#4);
        \addplot[forget plot,schem] table[col sep=space,x=f,y=mag]
          {figures/data/impedance/#1_meas.dat};
        \addplot[forget plot,schem,densely dashed] table[col sep=space,x=f,y=mag]
          {figures/data/impedance/#1_sim.dat};
      \nextgroupplot[
        height=0.90cm, ymin=-105, ymax=105, ytick={-90,0,90},
        ylabel={$\angle #2$ (\unit{\degree})}, xlabel={Frequency (\unit{\hertz})},
      ]
        \path[fill=figureBorder!8] (axis cs:1e6,-105) rectangle (axis cs:1e7,105);
        \addplot[forget plot,schem] table[col sep=space,x=f,y=phase]
          {figures/data/impedance/#1_meas.dat};
        \addplot[forget plot,schem,densely dashed] table[col sep=space,x=f,y=phase]
          {figures/data/impedance/#1_sim.dat};
    \end{groupplot}
    \node[anchor=north west,font=\footnotesize,fill=white,inner sep=1pt]
      at (group c1r1.north west) {\ValidationPanelLabel};
  \end{tikzpicture}%
}
\newcommand{\MaraPODOutputValidationFigure}{%
  \SingleTerminalValidationFigure{s2_zout}{\Zout}{1e-3}{1}{1e-3,1e-2,1e-1,1}%
}
\newcommand{\LvrInputValidationFigure}{%
  \SingleTerminalValidationFigure{s3_zin}{\Zin}{1e-2}{10}{1e-2,1e-1,1,10}%
}

\definecolor{nyquistOchre}{RGB}{160,117,28}
\definecolor{nyquistPurple}{RGB}{112,82,142}
\pgfplotsset{
  paper nyquist/.style={
    width=0.35\linewidth, height=2.0cm, scale only axis,
    xmajorgrids, ymajorgrids,
    grid style={draw=figureBorder!25, very thin},
    axis line style={draw=figureBorder}, tick style={draw=figureBorder},
    tick label style={font=\footnotesize, text=figureInk},
    label style={font=\footnotesize, text=figureInk},
    title style={font=\footnotesize, text=figureInk},
    every axis plot/.append style={line width=0.7pt, no marks},
    xlabel={$\operatorname{Re}\{\Tm\}$}, ylabel={$\operatorname{Im}\{\Tm\}$},
    legend style={draw=none, font=\footnotesize,
      /tikz/every even column/.append style={column sep=2pt}},
  },
}

\newcommand{\NyquistFrequencyLegend}{%
  \addlegendimage{empty legend}
  \addlegendentry{$f_{\min}=\qty{1}{\hertz}$}
  \addlegendimage{empty legend}
  \addlegendentry{$f_{\max}\approx\qty{1}{\mega\hertz}$}
}

\newcommand{\InterfaceOneNyquistFigure}{%
  \begin{tikzpicture}
    \begin{groupplot}[
      paper nyquist,
      group style={group size=2 by 1, horizontal sep=1.65cm},
    ]
    \nextgroupplot[
      xmin=-1.2, xmax=1.7, ymin=-1.15, ymax=0.6,
      xtick={-1,0,1}, ytick={-1,-0.5,0,0.5},
      legend to name=interfaceOneOverviewLegend, legend columns=9,
    ]
      \addlegendimage{empty legend}
      \addlegendentry{Cable length:}
      \addplot[figureBorder, densely dashed] table[col sep=comma,x=ReT_0m,y=ImT_0m]
        {figures/data/stability/chassis_nyq.csv};
      \addlegendentry{\qty{0}{\metre}}
      \addplot[schem] table[col sep=comma,x=ReT_80m,y=ImT_80m]
        {figures/data/stability/chassis_nyq.csv};
      \addlegendentry{\qty{80}{\metre}}
      \addplot[impedanceHeavy] table[col sep=comma,x=ReT_100m,y=ImT_100m]
        {figures/data/stability/chassis_nyq.csv};
      \addlegendentry{\qty{100}{\metre}}
      \addplot[nyquistOchre, densely dashdotted] table[col sep=comma,x=ReT_120m,y=ImT_120m]
        {figures/data/stability/chassis_nyq.csv};
      \addlegendentry{\qty{120}{\metre}}
      \addplot[nyquistPurple, dashed] table[col sep=comma,x=ReT_140m,y=ImT_140m]
        {figures/data/stability/chassis_nyq.csv};
      \addlegendentry{\qty{140}{\metre}}
      \addplot[schem, densely dotted, line width=0.9pt] table[col sep=comma,x=ReT_200m,y=ImT_200m]
        {figures/data/stability/chassis_nyq.csv};
      \addlegendentry{\qty{200}{\metre}}
      \NyquistFrequencyLegend
      \node[inner sep=0pt, text=figureInk] at (axis cs:-1,0) {$\times$};
      \node[anchor=south, font=\footnotesize, text=figureInk] at (axis cs:-1,0.06) {$-1$};
      \node[anchor=north west, font=\footnotesize, text=figureInk]
        at (rel axis cs:0.02,0.98) {(a)};
    \nextgroupplot[
      xmin=-0.27, xmax=0.19, ymin=-0.13, ymax=0.35,
      xtick={-0.2,0,0.1}, ytick={-0.1,0,0.1,0.2,0.3},
      scaled ticks=false, /pgf/number format/fixed,
    ]
      \addplot[schem] table[col sep=comma,x=ReT_80m,y=ImT_80m]
        {figures/data/stability/chassis_nyq.csv};
      \addplot[impedanceHeavy] table[col sep=comma,x=ReT_100m,y=ImT_100m]
        {figures/data/stability/chassis_nyq.csv};
      \addplot[nyquistOchre, densely dashdotted] table[col sep=comma,x=ReT_120m,y=ImT_120m]
        {figures/data/stability/chassis_nyq.csv};
      \addplot[nyquistPurple, dashed] table[col sep=comma,x=ReT_140m,y=ImT_140m]
        {figures/data/stability/chassis_nyq.csv};
      \addplot[schem, densely dotted, line width=0.9pt] table[col sep=comma,x=ReT_200m,y=ImT_200m]
        {figures/data/stability/chassis_nyq.csv};
      \node[anchor=north west, font=\footnotesize, text=figureInk]
        at (rel axis cs:0.02,0.98) {(b)};
    \end{groupplot}
  \end{tikzpicture}%
  \par\vspace{-3pt}%
  \pgfplotslegendfromname{interfaceOneOverviewLegend}%
}

\newcommand{\InterfaceTwoNyquistFigure}{%
  \begin{tikzpicture}
    \begin{groupplot}[
      paper nyquist,
      group style={group size=2 by 1, horizontal sep=1.65cm},
    ]
    \nextgroupplot[
      xmin=-5, xmax=125, ymin=-70, ymax=70,
      xtick={0,50,100}, ytick={-60,-30,0,30,60},
      legend to name=interfaceTwoOverviewLegend, legend columns=9,
    ]
      \addlegendimage{empty legend}
      \addlegendentry{Cable length:}
      \addplot[figureBorder, densely dashed] table[col sep=comma,x=ReT_0m,y=ImT_0m]
        {figures/data/stability/interface2_0pct_nyq_2T.csv};
      \addlegendentry{\qty{0}{\metre}}
      \addplot[nyquistOchre, densely dashdotted] table[col sep=comma,x=ReT_5m,y=ImT_5m]
        {figures/data/stability/interface2_0pct_nyq_2T.csv};
      \addlegendentry{\qty{5}{\metre}}
      \addplot[nyquistPurple, dashed] table[col sep=comma,x=ReT_10m,y=ImT_10m]
        {figures/data/stability/interface2_0pct_nyq_2T.csv};
      \addlegendentry{\qty{10}{\metre}}
      \addplot[schem] table[col sep=comma,x=ReT_20m,y=ImT_20m]
        {figures/data/stability/interface2_0pct_nyq_2T.csv};
      \addlegendentry{\qty{20}{\metre}}
      \addplot[impedanceHeavy] table[col sep=comma,x=ReT_30m,y=ImT_30m]
        {figures/data/stability/interface2_0pct_nyq_2T.csv};
      \addlegendentry{\qty{30}{\metre}}
      \addplot[schem, densely dotted, line width=0.9pt] table[col sep=comma,x=ReT_40m,y=ImT_40m]
        {figures/data/stability/interface2_0pct_nyq_2T.csv};
      \addlegendentry{\qty{40}{\metre}}
      \NyquistFrequencyLegend
      \node[anchor=north west, font=\footnotesize, text=figureInk]
        at (rel axis cs:0.02,0.98) {(c)};
    \nextgroupplot[
      xmin=-0.012, xmax=0.038, ymin=-0.024, ymax=0.017,
      xtick={-0.01,0,0.01,0.02,0.03}, ytick={-0.02,-0.01,0,0.01},
      scaled ticks=false, /pgf/number format/fixed,
    ]
      \addplot[schem] table[col sep=comma,x=ReT_20m,y=ImT_20m]
        {figures/data/stability/interface2_0pct_nyq_2T.csv};
      \addplot[impedanceHeavy] table[col sep=comma,x=ReT_30m,y=ImT_30m]
        {figures/data/stability/interface2_0pct_nyq_2T.csv};
      \addplot[schem, densely dotted, line width=0.9pt] table[col sep=comma,x=ReT_40m,y=ImT_40m]
        {figures/data/stability/interface2_0pct_nyq_2T.csv};
      \node[anchor=north west, font=\footnotesize, text=figureInk]
        at (rel axis cs:0.02,0.98) {(d)};
    \end{groupplot}
  \end{tikzpicture}%
  \par\vspace{-3pt}%
  \pgfplotslegendfromname{interfaceTwoOverviewLegend}%
}

\definecolor{inrushMag}{RGB}{31,72,125}
\definecolor{inrushPhase}{RGB}{171,52,40}
\newcommand{\InrushSimulationFigure}{%
      \begin{tikzpicture}
        \begin{axis}[
          width=0.80\linewidth, height=2.25cm, scale only axis,
          xmin=5, xmax=29, ymin=-0.4, ymax=4.7,
          xtick={5,15,25}, ytick={0,1,2,3,4},
          xmajorgrids, ymajorgrids,
          grid style={draw=figureBorder!25, very thin},
          axis line style={draw=figureBorder},
          tick style={draw=figureBorder},
          tick label style={font=\footnotesize, text=figureInk},
          label style={font=\footnotesize, text=figureInk},
          xlabel={Time (\unit{\milli\second})},
          ylabel={Input current (\unit{\ampere})},
        ]
          \addplot[draw=inrushMag, semithick, no marks]
            table[col sep=space, x=t_ms, y=I_A]
              {figures/data/inrush/inrush_sim.dat};
          \addplot[inrushMag, only marks, mark=*, mark size=1.4pt]
            coordinates {(17,4.217)};
          \node[font=\footnotesize, text=inrushMag, anchor=west]
            at (axis cs:18.0,4.217) {peak \qty{4.22}{\ampere}};
        \end{axis}
      \end{tikzpicture}%
}
\newcommand{\InrushMitigationFigure}{%
      \begin{tikzpicture}
        \begin{axis}[
          width=0.80\linewidth, height=2.25cm, scale only axis,
          xmin=0, xmax=1, ymin=0, ymax=1, enlargelimits=false,
          axis on top,
          xtick={0.135198,0.317716,0.500233,0.682751,0.865268},
          xticklabels={0,10,20,30,40},
          ytick={0.046809,0.436170,0.825532}, yticklabels={0,2,4},
          axis line style={draw=figureBorder},
          tick style={draw=figureBorder},
          tick label style={font=\footnotesize, text=figureInk},
          label style={font=\footnotesize, text=figureInk},
          xlabel={Time (\unit{\milli\second})},
          ylabel={Input current (\unit{\ampere})},
          /tikz/pk/.style={font=\footnotesize, fill=white, inner sep=0.6pt},
        ]
          \addplot graphics[xmin=0, xmax=1, ymin=0, ymax=1,
            includegraphics={trim=235bp 206bp 20bp 34bp,clip}]
            {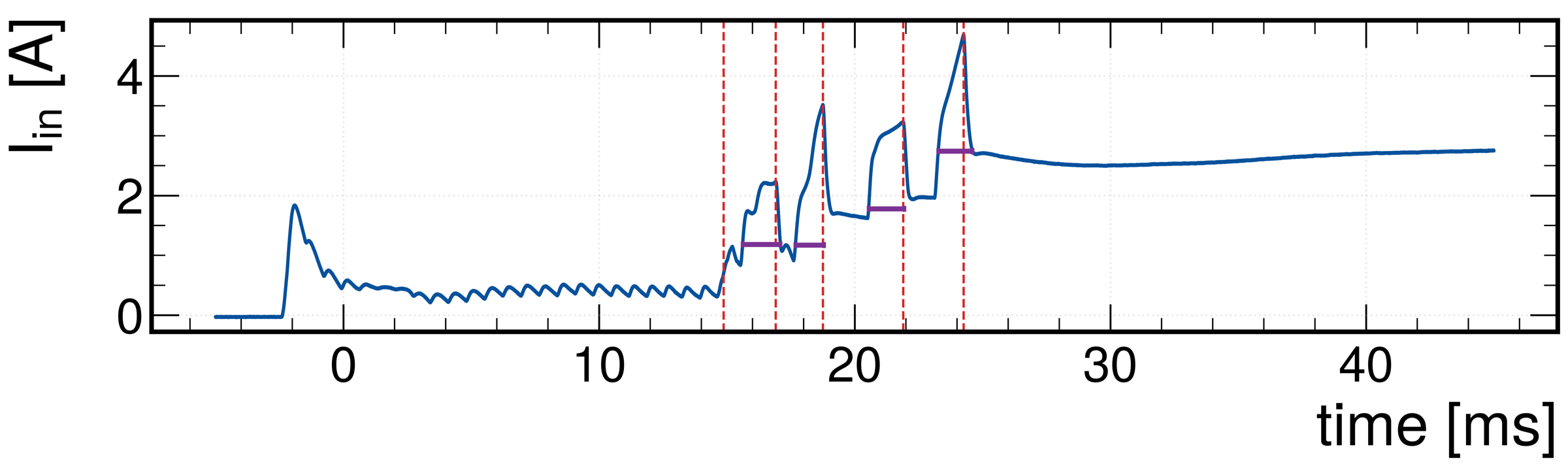};
          \node[pk, text=inrushMag, anchor=south east]
            at (axis cs:0.429745,0.505851) {A1};
          \node[pk, text=inrushMag, anchor=south east]
            at (axis cs:0.476737,0.755106) {B1};
          \node[pk, text=inrushPhase, anchor=south west]
            at (axis cs:0.548345,0.702234) {A2};
          \node[pk, text=inrushPhase, anchor=north east]
            at (axis cs:0.577436,0.989255) {B2};
        \end{axis}
      \end{tikzpicture}%
}

\DeclareSIUnit\dBohm{dB\ensuremath{\Omega}}
\DeclareSIUnit{\rad}{rad}        % radiation dose; use \mega\rad for "Mrad"

\newacro{HL-LHC}{High-Luminosity Large Hadron Collider}
\newacro{CMS}{Compact Muon Solenoid}
\newacro{ECAL}{electromagnetic calorimeter}
\newacro{LVR}{low-voltage regulator}
\newacro{PGOOD}{power-good}
\newacro{MLCC}{multilayer ceramic capacitor}
\newacro{IPA}{Institute for Particle Physics and Astrophysics}
\newacro{SNF}{Swiss National Science Foundation}
\newacro{AC}{alternating current}
\newacro{DC}{direct current}
\newacro{LC}{inductor-capacitor}
\newacro{RC}{resistor-capacitor}

\newcommand{\Zout}{Z_{\text{out}}}
\newcommand{\Zin}{Z_{\text{in}}}
\newcommand{\Tm}{T_{\mathrm{m}}}
\newcommand{\Nenc}{N_{\mathrm{enc}}}

\title{\acs*{CMS} \acs*{ECAL} Phase-2 upgrade: characterisation and stability analysis of a
  three-stage low-voltage power-conversion system}
\author[a,1]{Liv Anke Richter,\note{Corresponding author.}}
\author[b]{Günther Dissertori,}
\author[b]{Werner Lustermann}
\author[c]{and Krzysztof Stachon}
\affiliation[a]{ETH Zurich, Laboratory for High Power Electronic Systems,\\Zurich, Switzerland}
\affiliation[b]{ETH Zurich, Institute for Particle Physics and Astrophysics,\\Zurich, Switzerland}
\affiliation[c]{CERN,\\Geneva, Switzerland}
\emailAdd{lirichter@ethz.ch}
\proceeding{Topical Workshop on Electronics for Particle Physics (TWEPP 2026)\\
  28 September -- 2 October 2026\\
  Castelldefels, Barcelona, Spain}

\abstract{As part of the High-Luminosity upgrade of the Large Hadron Collider,
the front-end electronics of the \acl*{CMS} \acl*{ECAL}
barrel are being replaced. A three-stage low-voltage power-conversion system supplies the
upgraded electronics. As the cascade dynamics require characterisation to ensure
stable operation, interactions between the three converter stages are assessed
at their two interfaces using the Nyquist criterion. A validated shunt-injection
setup using a Bode~100 frequency-response analyser provides in-situ impedance
measurements under load to assess the accuracy of the averaged converter models.
Results show that both interfaces are stable over the
evaluated configurations, with modelled minimum distances to the critical point
of $0.76$ and $0.79$, respectively. A turn-on study identifies capacitor
charging at Stage~3 as the main cause of inrush currents triggering
Stage~2 overcurrent protection. Two Stage~3 board variants stagger converter
enables within and between boards, keeping the measured two-tower current below
\qty{5}{\ampere}, within the \qty{8.0}{\ampere} protection threshold.}


\begin{document}
\maketitle
\clearpage
\setcounter{page}{1}
\acresetall
\flushbottom

\section{Introduction}
\label{sec:intro}
For operation at the \ac{HL-LHC}, the \ac{CMS}~\cite{CMS_Experiment,CMS_Run3}
\ac{ECAL}~\cite{ECAL_TDR} barrel receives new front-end
electronics. The upgraded readout electronics~\cite{CMS_Phase2_TDR} require supply
rails at \qty{1.2}{\volt} and \qty{2.5}{\volt}~\cite{gadek_dcdc}. Distributing these
low voltages directly from remote supplies over long cables would result in prohibitive
resistive losses, $P_{\mathrm{loss}}=I^2R$. Point-of-load conversion at the front-end
is therefore adopted for the Phase-2 upgrade.

The resulting power-distribution system comprises three cascaded conversion stages.
Although each stage is individually stable, interactions between their terminal
impedances can destabilise the cascade~\cite{middlebrook}. This work assesses
stability at both stage interfaces using averaged small-signal models and in-situ
impedance measurements.
Moreover, turn-on inrush current is investigated, and a mitigation
based on staggered converter enables is evaluated.

\section{Materials and Methods}
\label{sec:methods}
\subsection{System and stability criterion}
\label{sec:system}
\begin{figure}[htbp]
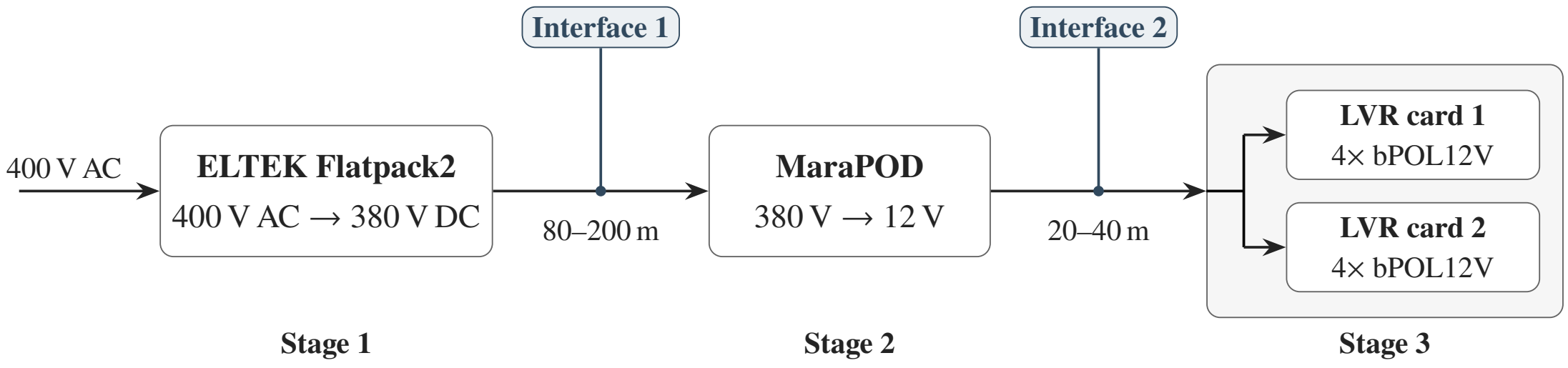

  \centering
  \begin{minipage}{0.96\textwidth}
    \PowerChainFigure
  \end{minipage}
  \caption{\label{fig:chain}Three-stage \acs*{ECAL} barrel power distribution with Interfaces~1 and~2.}
\end{figure}

The power-distribution system consists of an \acs*{AC}--\acs*{DC} rectifier followed by two \acs*{DC}--\acs*{DC}
conversion stages, as shown in figure~\ref{fig:chain}.
Stage~1, the ELTEK Flatpack2, supplies a \qty{380}{\volt} \acs*{DC}
bus~\cite{flatpack2_datasheet}. Stage~2, the Kontron MaraPOD,
provides a \qty{12}{\volt} \acs*{DC} output. Stage~3 \ac{LVR} cards implement
bPOL12V \acs*{DC}--\acs*{DC} converter-based point of load
conversion~\cite{gadek_dcdc,bpol12v_datasheet}. Each \ac{LVR} card supplies
two \qty{1.2}{\volt} and two \qty{2.5}{\volt} rails to one so-called readout
tower, serving 25 readout channels.
Two \ac{LVR} cards are supplied by a single MaraPOD channel. Interface~1 denotes the
Flatpack2--MaraPOD connection, and Interface~2 the MaraPOD--\ac{LVR} connection.
The connecting cables at both interfaces are also considered over a range of
lengths in this work.

A regulated downstream converter supplying constant output power behaves
approximately as a constant-power load within its regulation bandwidth. A fall in
input voltage therefore increases its input current, corresponding to a negative
incremental input impedance. This additional current causes a further voltage drop
across the finite output impedance of the upstream converter. Depending on the
frequency-dependent impedances of both stages, this interaction can amplify
disturbances and lead to oscillations. The interstage interaction can be represented
as a feedback loop, whose small-signal interface-voltage response
is~\cite{middlebrook},
\begin{equation}
  \label{eq:minor-loop}
  H(s)=\frac{\hat v_{\mathrm{in}}(s)}{\hat v_s(s)}
    =\frac{1}{1+\Tm(s)},
  \qquad
  \Tm(s)=\frac{\Zout(s)}{\Zin(s)},
\end{equation}
where $\hat v_s$ is the undisturbed source-voltage perturbation and
$\hat v_{\mathrm{in}}$ the interface-voltage perturbation. The source output
impedance $\Zout$ and load input impedance $\Zin$ thus determine the interstage
loop gain $\Tm$, called the minor-loop gain.

Middlebrook's sufficient condition, $|\Zout|\ll|\Zin|$ at all frequencies, keeps
the minor-loop gain small but is conservative because it disregards phase.
The interface transfer function $H(s)$ becomes
singular at $\Tm=-1$, where the denominator $1+\Tm$ vanishes. This requires both
unit loop-gain magnitude and a \ang{180} phase shift. Phase must therefore be
included in the stability assessment. Violating the magnitude condition alone
does not establish instability.

The Nyquist plot traces $\Tm(j\omega)$ in the complex plane, retaining magnitude
and phase. For a clockwise right-half-plane contour, Cauchy's
argument principle~\cite{brown_churchill} gives $\Nenc=Z-P$, where $\Nenc$ is
the net number of clockwise encirclements of $-1$ by the complete
$\Tm$ locus. Here $P$ counts right-half-plane poles of $\Tm$, and $Z$ counts
right-half-plane zeros of $1+\Tm$, corresponding to unstable poles of
$H(s)$. The source and load stages are individually stable. Assuming
no hidden unstable pole--zero cancellations, $P=0$ and hence $\Nenc=Z$. Stability
therefore requires $\Nenc=0$. The Nyquist plot additionally gives a direct measure of the separation
from $-1$: the shortest distance $\min|1+\Tm|$ over the evaluated frequencies,
known as the vector margin.

\subsection{Modelling for small-signal analysis}
\label{sec:model}
One approach used to obtain the terminal impedances is to model the converters in LTspice.
For Stage~1, a behavioural terminal model comprising a passive \acs*{LC} output filter
and a regulation loop was constructed and fitted to the measured output impedance.

The regulated output stage of Stage~2 and the converters of Stage~3 are synchronous
buck converters. Averaging over a switching period removes the explicit time
dependence introduced by switching while retaining the slower filter and
control-loop dynamics. Canonical averaged switch
models~\cite{erickson} were implemented in total-signal form: the
behavioural sources act on the full averaged voltages, currents and duty ratio.
Small-signal perturbations are injected at the model terminals. The resulting
voltage-to-current ratios give the terminal impedances.

Average models lose accuracy at higher frequencies. The crossover limits of
\qty{410}{\hertz} and \qty{31}{\kilo\hertz} lie well below the validity limits
($f_s/5$), corresponding to \qty{34}{\kilo\hertz} and \qty{360}{\kilo\hertz},
for Stages~2 and~3, respectively.
Averaged modelling therefore describes the control-loop dynamics and
low-frequency constant-power-load behaviour with sufficient accuracy.
Where propagation effects were relevant, cables were represented by distributed
transmission-line models.

\subsection{Impedance measurement}
\label{sec:meas}
As a second approach, impedances were measured directly by shunt injection
(figure~\ref{fig:meas}).
A Bode~100 analyser~\cite{omicron_bode100} drives a B-WIT~100 injection
transformer~\cite{omicron_bwit100}, coupling a small \acs*{AC} perturbation in parallel
with the interface through a \acs*{DC} block. The converters remain connected and operate
under load. The sensed \acs*{AC} voltage and branch current give the complex impedance,
\mbox{$Z(j\omega)=V(j\omega)/I(j\omega)$}. The current-clamp position selects the measured
impedance: the source feed gives $\Zout$, while the load branch gives $\Zin$.
Reference-load checks reproduced the known impedance in the \unit{\milli\ohm} range above
approximately \qty{100}{\hertz}. Lower-frequency data were excluded because of
\qty{50}{\hertz} interference.

\subsection{Turn-on inrush}
\label{sec:inrush-methods}
Stage~3 input current was measured at the \qty{12}{\volt} bus with an oscilloscope.
The four converters on each \ac{LVR} card start in two \ac{PGOOD}-controlled groups:
C/D, then A/B, with about \qty{520}{\micro\second} internal soft-start
per converter~\cite{bpol12v_datasheet}.
The averaged switch models include soft-start, \ac{PGOOD} sequencing and the
A/B load capacitances of about \qty{1.35}{\milli\farad} and
\qty{1.28}{\milli\farad}, respectively, including very-front-end decoupling
and \ac{LVR} output filters.

The LVR5A and LVR5B cards use different enable-pin \acs*{RC} delays to stagger
A/B start-up within and between the two cards sharing a MaraPOD channel.
Table~\ref{tab:enable-rc} lists their A/B enable capacitances $C_{\mathrm{En}}$
and resistances $R_{\mathrm{En}}$. C/D use \qty{4.7}{\micro\farad} on both variants.

% RC values
\begingroup
\nolinenumbers
\captionsetup{hypcap=false}
\par\addvspace{\intextsep}
\noindent
  \begin{minipage}[c]{0.46\textwidth}
    \centering
    \makebox[\linewidth][c]{\begin{minipage}{0.90\linewidth}\MeasSetupFigure\end{minipage}}
    \captionof{figure}{\label{fig:meas}Shunt-injection measurement of the MaraPOD output impedance $\Zout$ at Interface~2.}
  \end{minipage}\hfill
  \begin{minipage}[c]{0.50\textwidth}
    \centering
    \small
    \captionof{table}{\label{tab:enable-rc}Enable-network configurations for the two \acs*{LVR}-card variants.}
    \setlength{\tabcolsep}{3pt}
    \begin{tabular}{lcccc}
      \toprule
      & \multicolumn{2}{c}{Converter A} & \multicolumn{2}{c}{Converter B} \\
      Card variant & $C_{\mathrm{En}}$ & $R_{\mathrm{En}}$ & $C_{\mathrm{En}}$ & $R_{\mathrm{En}}$ \\
      \midrule
      LVR5A & \qty{0.47}{\micro\farad} & \qty{1}{\kilo\ohm}
            & \qty{2.2}{\micro\farad} & \qty{1}{\kilo\ohm} \\
      LVR5B & \qty{4.7}{\micro\farad} & \qty{1}{\kilo\ohm}
            & \qty{4.7}{\micro\farad} & \qty{1.47}{\kilo\ohm} \\
      \bottomrule
    \end{tabular}
  \end{minipage}
\par\addvspace{\intextsep}
\endgroup

\raggedbottom % Avoid stretched heading gaps when a result figure starts the next page.
\section{Results}
\label{sec:results}
\begingroup

\subsection{Model validation}
\label{sec:validation-results}
Figure~\ref{fig:val-group} compares measured (solid) and modelled
(dashed) terminal impedances. The overlays check model accuracy for the subsequent
stability analysis.

\begin{figure}[H]
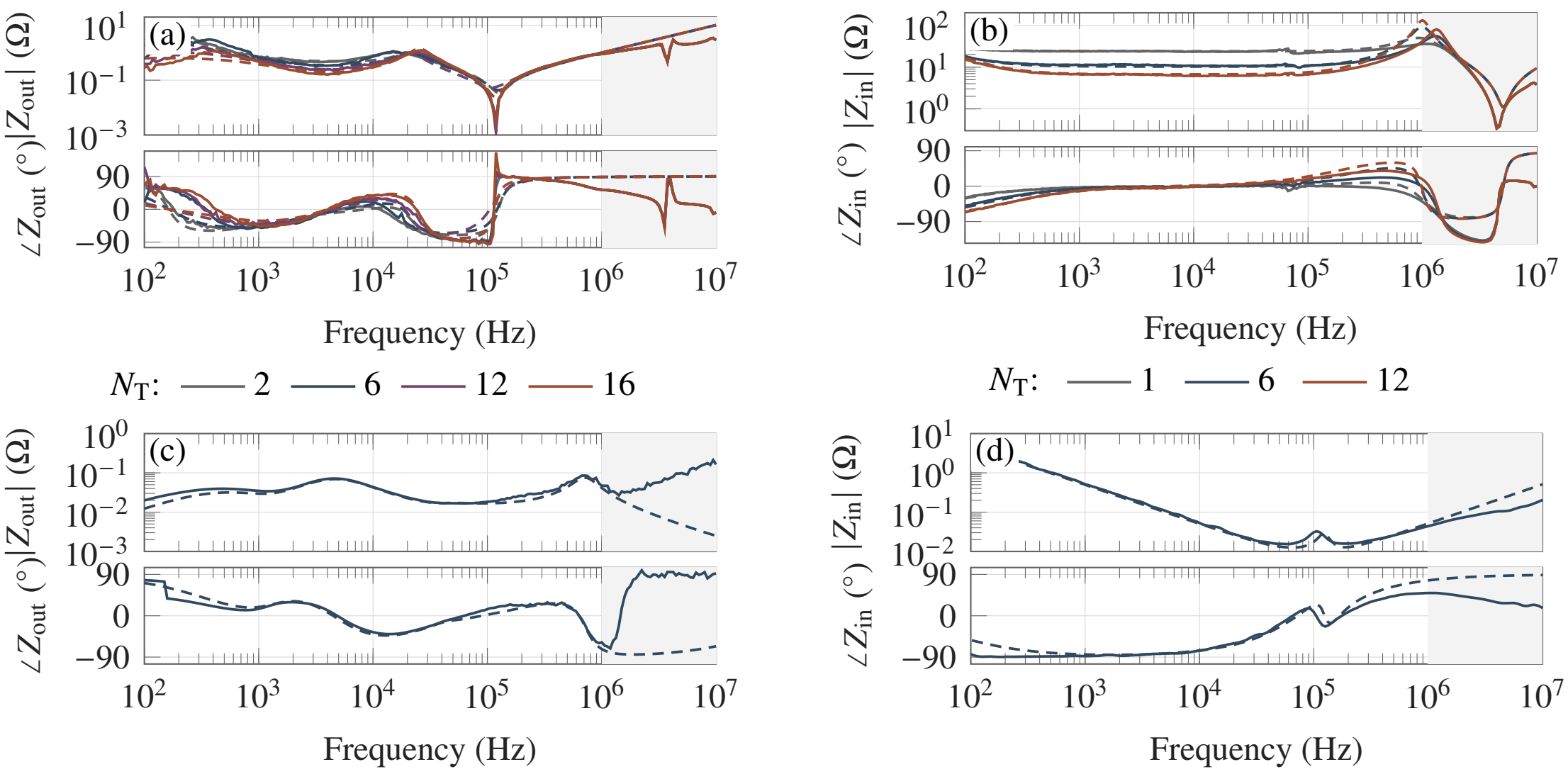

  \centering
  \begin{subfigure}[t]{0.49\textwidth}
    \centering\vspace{0pt}
    \phantomsubcaption\label{fig:val-eltek}
    \renewcommand{\ValidationPanelLabel}{(a)}
    \EltekImpedanceValidationFigure
  \end{subfigure}\hfill
  \begin{subfigure}[t]{0.49\textwidth}
    \centering\vspace{0pt}
    \phantomsubcaption\label{fig:val}
    \renewcommand{\ValidationPanelLabel}{(b)}
    \ImpedanceOverlayFigure
  \end{subfigure}

  \vspace{-4pt}
  \begin{subfigure}[t]{0.49\textwidth}
    \centering\vspace{0pt}
    \phantomsubcaption\label{fig:val-if2-out}
    \renewcommand{\ValidationPanelLabel}{(c)}
    \MaraPODOutputValidationFigure
  \end{subfigure}\hfill
  \begin{subfigure}[t]{0.49\textwidth}
    \centering\vspace{0pt}
    \phantomsubcaption\label{fig:val-if2-in}
    \renewcommand{\ValidationPanelLabel}{(d)}
    \LvrInputValidationFigure
  \end{subfigure}
  \caption{\label{fig:val-group}Terminal impedances: (a) ELTEK output and
    (b) MaraPOD input at Interface~1; (c) MaraPOD output and (d) \acs*{LVR} input
    at Interface~2 ($N_{\mathrm{T}}=1$).}
\end{figure}

All four comparisons show deviations above approximately \qty{1}{\mega\hertz}
(shaded region). These deviations may reflect unmodelled parasitic effects and
reduced model accuracy. For the MaraPOD output measurement, insufficient
signal-to-noise ratio limits agreement at higher frequencies. Legends indicate
the energised tower count $N_{\mathrm{T}}$.

\subsection{Small-signal stability}
\label{sec:stability-results}
The Nyquist plots shown for both interfaces are based on simulated terminal impedances.
At Interface~1, with one MaraPOD module connected, $|\Tm|\leq0.24$ and $\min|1+\Tm|\geq0.95$.
Ten paralleled modules in a full chassis reduce their combined input impedance
to $\Zin/10$. At short cable lengths, the Middlebrook magnitude condition is
locally violated, but we found $\Nenc=0$ at every evaluated length
(figure~\ref{fig:nyquist-if1}a). The closest approach, $\min|1+\Tm|=0.76$, occurs
with a \qty{100}{\metre} cable near \qty{9}{\hertz}, below the measurement floor,
and is therefore model-based. Independent \qty{\pm50}{\percent} variations of
the fitted Stage~1 loop resistance and capacitance preserve $\Nenc=0$, with
minimum distances of approximately $0.76$.

Additional parallel chassis move the modelled loci away from $-1$.

\smallskip
At Interface~2, for one readout tower, the ratio of the measured MaraPOD output and \ac{LVR} input
impedances gives no encirclement of $-1$ over the available band, with
$\min|1+\Tm|=0.93$.
The nominal two-tower assessment is model-based, paralleling two identical
\ac{LVR} input models validated at one tower.

The two-tower sweep over cable lengths of \qtyrange{0}{40}{\metre} and effective
capacitances of \qtylist{20;40;60;88}{\micro\farad} per \ac{MLCC} output bank gives $\Nenc=0$ throughout. Figure~\ref{fig:nyquist-if2}(c,d) shows
\qty{0}{\percent} derating (\qty{88}{\micro\farad} per bank), which gives the
smallest margin among the evaluated capacitances: $\min|1+\Tm|=0.79$ without
a cable. With \qtyrange{20}{40}{\metre} service cables, the load impedance exceeds
the source by \qtyrange{30}{36}{\decibel} and $\min|1+\Tm|=0.995$--$0.998$.
Cable damping increases the stability margin.

\begin{figure}[H]
  \centering
  \InterfaceOneNyquistFigure

  \vspace{-2pt}
  \InterfaceTwoNyquistFigure
  \caption{\label{fig:nyquist-group}\label{fig:nyquist-if1}\label{fig:nyquist-if2}Modelled Nyquist loci:
    (a,b) Interface~1, full chassis, \qtyrange[range-phrase=--,range-units=single]{0}{200}{\metre}
    (zoom: \qtyrange[range-phrase=--,range-units=single]{80}{200}{\metre});
    (c,d) Interface~2, two towers, \qty{0}{\percent} derating,
    \qtyrange[range-phrase=--,range-units=single]{0}{40}{\metre}
    (zoom: \qtyrange[range-phrase=--,range-units=single]{20}{40}{\metre}).}
\end{figure}
\endgroup

\newpage % Keep the turn-on results together after the grouped Nyquist figure.
\subsection{Turn-on inrush}
\label{sec:inrush-results}
Oscilloscope measurements showed that the inrush
current on the \qty{12}{\volt} supply exceeds \qty{4}{\ampere} per tower.
When both towers are enabled, their combined inrush current exceeds the MaraPOD's
\qty{8.0}{\ampere} overcurrent threshold and causes it to latch off.

Converters A and B dominate the measured inrush current. For a single LVR card,
the averaged model predicts a peak input current of \qty{4.22}{\ampere} (figure~\ref{fig:inrush}).

With the staggered enable configurations listed in table~\ref{tab:enable-rc},
the measured two-tower current remains below \qty{5}{\ampere}
(figure~\ref{fig:inrush-mitigated}), even when both cards are powered together.

\begingroup
\begin{figure}[H]
  \centering
  \begin{minipage}[t]{0.49\textwidth}
    \centering
    \vspace{0pt}
    \parbox[c][3.2cm][c]{\linewidth}{\centering\InrushSimulationFigure}
    \caption{\label{fig:inrush}Simulated input current at the \qty{12}{\volt} bus during single-tower turn-on.}
  \end{minipage}\hfill
  \begin{minipage}[t]{0.49\textwidth}
    \centering
    \vspace{0pt}
    \parbox[c][3.2cm][c]{\linewidth}{\centering\InrushMitigationFigure}
    \caption{\label{fig:inrush-mitigated}Measured two-tower turn-on with staggered
      enables. A/B: converters, 1/2: \acs*{LVR} cards.}
  \end{minipage}
\end{figure}
\endgroup

\section{Discussion and conclusion}
\label{sec:concl}
Averaged converter models and in-situ impedance measurements were used to assess
the small-signal stability of the three-stage \ac{ECAL} power-distribution system.
The modelled terminal impedances show good agreement with measurements over
the considered frequency range, up to approximately \qty{1}{\mega\hertz}.
For all evaluated configurations, the modelled Nyquist loci show no encirclement
of $-1$, including cases that violate the Middlebrook magnitude condition.
Both interfaces remain stable, with modelled minimum
distances to the critical point, $\min|1+\Tm|$, of $0.76$ at Interface~1
and $0.79$ at Interface~2. These conclusions apply to small-signal
operation and rely partly on fitted converter models.

The turn-on current measurements identify converters A and B charging their capacitive
loads as the main contributors to the Stage~3 inrush current. Staggering their
enables keeps the measured two-tower current below \qty{5}{\ampere}, within the
MaraPOD's \qty{8.0}{\ampere} latch-off threshold, preventing trips.

\clearpage
\acknowledgments
The authors thank Berkan Kaynak for providing the MaraPOD setup and Ashley
Churchmann and Dominik Freyberg for their practical support throughout this work.
The authors thank Marcel Reiff from Kontron HARTMANN-WIENER for his technical
support and helpful discussions.
This research was supported by the \ac{SNF} via the grant
\acs*{SNF} FLARE 20FL20\_232672.

\bibliographystyle{JHEP}
\bibliography{refs}

\end{document}